\documentstyle[12pt]{article}
\def\journal{\topmargin .3in    \oddsidemargin .5in
        \headheight 0pt \headsep 0pt
        \textwidth 5.625in 
        \textheight 8.25in 
        \marginparwidth 1.5in
        \parindent 2em
        \parskip .5ex plus .1ex         \jot = 1.5ex}
\journal

\catcode`\@=11
\def\marginnote#1{}
\catcode`\@=11
\def\section{\@startsection {section}{1}{0pt}{-3.5ex plus -1ex minus
 -.2ex}{2.3ex plus .2ex}{\raggedright\large\bf}}
\catcode`\@=12
\newskip\humongous \humongous=0pt plus 1000pt minus 1000pt
\def\caja{\mathsurround=0pt}

\newif\ifdtup
\def\panorama{\global\dtuptrue \openup1\jot \caja
        \everycr{\noalign{\ifdtup \global\dtupfalse
        \vskip-\lineskiplimit \vskip\normallineskiplimit
        \else \penalty\interdisplaylinepenalty \fi}}}
\def\eqalignno#1{\panorama \tabskip=\humongous
        \halign to\displaywidth{\hfil$\displaystyle{##}$
        \tabskip=0pt&$\displaystyle{{}##}$\hfil
        \tabskip=\humongous&\llap{$##$}\tabskip=0pt
        \crcr#1\crcr}}

\def\one{{\mathchoice {\rm 1\mskip-4mu l} {\rm 1\mskip-4mu}
{\rm 1\mskip-4.5mu l} {\rm 1\mskip-5mu l}}}
\def\Q{{\mathchoice
{\setbox0=\hbox{$\displaystyle\rm Q$}\hbox{\raise 0.15\ht0\hbox to0pt
{\kern0.4\wd0\vrule height0.8\ht0\hss}\box0}}
{\setbox0=\hbox{$\textstyle\rm Q$}\hbox{\raise 0.15\ht0\hbox to0pt
{\kern0.4\wd0\vrule height0.8\ht0\hss}\box0}}
{\setbox0=\hbox{$\scriptstyle\rm Q$}\hbox{\raise 0.15\ht0\hbox to0pt
{\kern0.4\wd0\vrule height0.7\ht0\hss}\box0}}
{\setbox0=\hbox{$\scriptscriptstyle\rm Q$}\hbox{\raise 0.15\ht0\hbox to0pt
{\kern0.4\wd0\vrule height0.7\ht0\hss}\box0}}}}
\def\C{{\mathchoice
{\setbox0=\hbox{$\displaystyle\rm C$}\hbox{\hbox to0pt
{\kern0.4\wd0\vrule height0.9\ht0\hss}\box0}}
{\setbox0=\hbox{$\textstyle\rm C$}\hbox{\hbox to0pt
{\kern0.4\wd0\vrule height0.9\ht0\hss}\box0}}
{\setbox0=\hbox{$\scriptstyle\rm C$}\hbox{\hbox to0pt
{\kern0.4\wd0\vrule height0.9\ht0\hss}\box0}}
{\setbox0=\hbox{$\scriptscriptstyle\rm C$}\hbox{\hbox to0pt
{\kern0.4\wd0\vrule height0.9\ht0\hss}\box0}}}}

\font\fivesans=cmss10 at 4.61pt
\font\sevensans=cmss10 at 6.81pt
\font\tensans=cmss10
\newfam\sansfam
\textfont\sansfam=\tensans\scriptfont\sansfam=\sevensans\scriptscriptfont
\sansfam=\fivesans
\def\sans{\fam\sansfam\tensans}
\def\Z{{\mathchoice
{\hbox{$\sans\textstyle Z\kern-0.4em Z$}}
{\hbox{$\sans\textstyle Z\kern-0.4em Z$}}
{\hbox{$\sans\scriptstyle Z\kern-0.3em Z$}}
{\hbox{$\sans\scriptscriptstyle Z\kern-0.2em Z$}}}}

\mathchardef\endbar="375

\def\ceilfill{$\raise3pt\hbox{$\mathsurround=0pt\mathord\endbar$}
  \mkern-2mu \xleaders\hbox{$\mkern-5mu
  \mathord-\mkern-5mu$}\hfill\mkern-7mu
  \raise3pt\hbox{$\mathsurround=0pt\mathord\endbar$}$}

\def\floorfill{$\raise9pt\hbox{$\mathsurround=0pt\mathord\endbar$}
  \mkern-2mu \xleaders\hbox{$\mkern-5mu
  \mathord-\mkern-5mu$}\hfill\mkern-7mu
  \raise9pt\hbox{$\mathsurround=0pt\mathord\endbar$}$}

\def\overcontract#1{\mathop{\vbox{\ialign{##\crcr\noalign{\kern3pt}
  \ceilfill\hskip6pt\crcr\noalign{\kern3pt\nointerlineskip}
  $\hfil\displaystyle{#1}\hfil$\crcr}}}}

\def\undercontract#1{\mathop{\vtop{\ialign{##\crcr
  $\hfil\displaystyle{#1}\hfil$\crcr\noalign{\kern3pt\nointerlineskip}
  \floorfill\hskip6pt\crcr\noalign{\kern3pt}}}}}
\def\a{\alpha}
\def\b{\beta}
\def\g{\gamma}
\def\d{\delta}

\def\n{\nu}

\def\p{\pi}
\def\ps{\psi}

\def\o{\omega}

\def\et{\eta}

\def\D{\Delta}
\def\bz{\bar{z}}

\def\bu{\bar{u}}

\def\bps{\bar{\psi}}

\def\bA{\bar{A}}

\def\bY{\bar{Y}}

\def\tL{\tilde{L}}
\def\tT{\tilde{T}}

\def\F{{\cal F}}

\def\T{{\cal T}}
\def\C{{\cal C}}
\def\cO{{\cal O}}

\def\pa{\partial}
\def\bpa{\bar{\partial}}
\def\ve{\vert}

\def\ra{\rightarrow}
\def\lra{\leftrightarrow}

\def\ci{\cite}
\def\xx{\hbox{ }^*_*}

\def\ref#1{$^{#1)}$}

\begin{document}
\begin{titlepage}
\begin{center}
 September, 1993        \hfill     UCB-PTH-93/25 \\
hep-th/9309087    \hfill     LBL-34610 \\
\vskip .5in

{\large \bf Recent Progress in  Irrational Conformal Field Theory }
\footnote{This work was supported in part by the Director, Office of
Energy Research, Office of High Energy and Nuclear Physics, Division of
High Energy Physics of the U.S. Department of Energy under Contract
DE-AC03-76SF00098 and in part by the National Science Foundation under
grant PHY90-21139.}
\vskip .3in
M.B. Halpern
\footnote{e-mail: MBHALPERN@LBL.GOV, THEORY::HALPERN}
\footnote{Talk presented at the conference ``Strings 1993'', Berkeley,
 May 23-29.}
\vskip .2in
{\em  Department of Physics \\ University of California\\ and \\
      Theoretical Physics Group \\ Physics Division \\
     Lawrence Berkeley Laboratory\\ 1 Cyclotron Road \\
      Berkeley, California 94720 \\
       USA}

\end{center}
\vskip .3in
\begin{abstract}
In this talk, I will review the foundations of irrational conformal field
theory (ICFT), which includes rational conformal field theory as a small
subspace. Highlights of the review include the Virasoro master equation,
the Ward identities for the correlators of ICFT and solutions of the Ward
identities. In particular, I will discuss the solutions for the correlators
of the $g/h$ coset constructions and the correlators of the affine-Sugawara
nests on $g\supset h_1 \supset \ldots \supset h_n$. Finally, I will
discuss the recent global solution for the correlators of all the ICFT's
in the master equation.

\end{abstract}

\end{titlepage}
\renewcommand{\thepage}{\arabic{page}}
\setcounter{page}{1}
\setcounter{footnote}{1}

\setcounter{section}{-1}
\section{Outline of the Talk}
\begin{enumerate}
\item History of the affine-Virasoro constructions.
\item The general affine-Virasoro construction.
\item Irrational conformal field theory.
\item The Ward identities of irrational CFT.
\item Coset and nest correlators.
\item Algebraization of the Ward identities.
\item Candidate correlators for irrational CFT.
\item Conclusions.
\end{enumerate}

\section{History of the Affine-Virasoro Constructions}
 Affine-Virasoro constructions are Virasoro operators constructed with the
currents $J_a,\, a=1 , \ldots,\mbox{dim}\,g$ of affine Lie $g$. All known
conformal field theories may be constructed this way.

Here is a brief history of these constructions, which began with two\break
papers~[1,2]
by Bardakci and myself in 1971. These papers contained the following
developments.

\noindent a) The first examples of affine Lie algebra, or current algebra on
$S^1$. This was the independent discovery of affine Lie algebra in physics,
including the
\hbox{affine central extension some years before it was
recognized in mathematics~\ci{lw}.}
 Following the convention in math,
we prefer the neutral names affine
Lie
algebra or
current algebra, reserving ``Kac-Moody'' for the more general case~\ci{km}
including hyperbolic algebras.

\noindent b) World-sheet fermions (half-integer moded), from which the affine
algebras were constructed. Ramond \ci{ram}
gave the integer-moded case in the same issue of Physical Review.

\noindent c) The first affine-Sugawara constructions, on the currents of affine
Lie algebra. Sugawara's model \ci{sug} was in four dimensions on a
different algebra. The
affine-Sugawara
constructions were later generalized by Knizhnik and
Zamolodchikov~\ci{kz}
and
Segal~\cite{se}, and the corresponding world-sheet action was given
by \hbox{Witten}~\ci{wit1}.

\noindent d) The first coset constructions, implicit in [1] and
explicit in [2], which were later generalized by Goddard, Kent and
Olive [10].

What was forgotten for many years was that our first paper also gave another
affine-Virasoro
construction, the so-called ``spin-orbit'' model, which was more general than
the affine-Sugawara and coset constructions.

\section{The General Affine-Virasoro Construction}
Motivated by the affine-Sugawara, coset and spin-orbit constructions, Kiritsis
and I studied the general affine-Virasoro construction [11,12] in 1989.

The construction begins with the currents of untwisted affine
Lie $g $  [4,1]
$$J_a(z)\, J_b(w)={G_{ab} \over (z-w)^2}+ { i{f_{ab}}^c \over z-w} \, J_c(w) +
\cO (z-w)^0 \eqno(2.1) $$
where $a,b = 1,\ldots, \mbox{dim}\,g$ and ${f_{ab}}^c$ and $G_{ab}$ are
respectively the structure constants and
generalized Killing metric of
$g= \oplus_I g_I $. To obtain level
$x_I = 2k_I/\psi_I^2$ of affine $g_I$ with dual Coxeter number
$\tilde{h}_I= Q_I /\psi_I^2$, take
$$G_{ab}=\oplus_I k_I \, \et_{ab}^I\;\;\;\;,\;\;\;\;
f_{ac}{}^d f_{bd}{}^c = - \oplus_I Q_I \, \eta_{ab}^I
\eqno(2.2) $$
where $\eta_{ab}^I$ and $\psi_I$ are respectively the Killing metric and
 highest root of $g_I$.

Given the currents, we consider the general stress tensor
$$ T= L^{ab} \xx J_a J_b \xx + D^a \partial J_a + d^a J_a
\eqno(2.3) $$
where the coefficients $L^{ab}= L^{ba}$, $D^a$ and $d^a$ are to be determined.
$L^{ab}$ is called the inverse inertia tensor, in analogy with the spinning
top. The stress tensor is required to satisfy the Virasoro algebra
$$T(z)\,T(w)={c/2 \over (z-w)^4} +\left(\frac{2}{(z-w)^2}+{ \pa_w \over z-w}
\right)T(w)+ \cO (z-w)^0 \eqno(2.4)$$
where $c$ is the central charge. I give here only the result at $D^a=d^a=0$,
which is called the Virasoro master equation [11,12]
$$ L^{ab}=2 L^{ac}G_{cd}L^{db}-L^{cd}L^{ef}{f_{ce}}^{a}{f_{df}}^{b}-
L^{cd}{f_{ce}}^{f}{f_{df}}^{(a}L^{b)e} \eqno(2.5a)$$
$$ c=2 \,G_{ab}L^{ab}\;\;\;\;.\eqno(2.5b)$$
The master equation has been identified \ci{hy1} as an Einstein
system on the group manifold, with
$$ g^{ij}= e_a^i L^{ab} e_b^j \;\;\;\;,\;\;\;\;
c=\mbox{dim}\,g-4R \eqno(2.6)$$
where $g^{ij}$ and $R$  are the inverse Einstein metric
and  Einstein curvature scalar respectively.

Here are some simple facts about the master equation which we will need in
this lecture.

\noindent a) Affine-Sugawara constructions [1,2,7-9]. The
affine-Sugawara construction on $g$ is
$$ L_g^{ab}=\oplus_I {\eta_I^{ab} \over 2k_I +Q_I}\;\;\;\,,\;\;\;\;\;\;\;
c_g=\sum_I {x_I \, \mbox{dim}\,g_I \over x_I + \tilde{h}_I } \eqno(2.7) $$
and similarly for $L_h$ on $h \subset g$.

\noindent b) K-conjugation covariance [1,2,10,11]. Solutions of the
master equation come in commuting K-conjugate pairs
$T=L^{ab} \xx J_a J_b \xx$ and
$\tilde{T}=\tilde{L}^{ab} \xx J_a J_b \xx$, which sum to the affine-Sugawara
construction $T_g=L_g^{ab} \xx J_a J_b \xx$,
$$ L + \tilde{L}=L_g \;\;\;,\;\;\;\;\;\;\; T + \tilde{T} = T_g
\;\;\;\;,\;\;\;\;
c + \tilde{c}=c_g
\eqno(2.8a) $$
$$ T(z)\, \tilde{T} (w)= \cO (z-w)^0  \;\;\;\;\;\; . \eqno(2.8b) $$
K-conjugation is the central feature of affine-Virasoro constructions, and it
suggests that the affine-Sugawara construction should be thought of as the
tensor product of any pair of K-conjugate conformal field theories.
This is the conceptual basis of factorization, discussed in Sections 4-7.

\noindent c) Coset constructions [1,2,10].
K-conjugation generates new solutions from
old. The simplest examples are the $g/h$ coset constructions
$$ \tilde{L} =  L_g - L_h = L_{g/h} \;\;\;,\;\;\;\;
\tilde{T} =  T_g - T_h = T_{g/h}
\;\;\;,\;\;\;\; \tilde{c} = c_g-c_h = c_{g/h} \eqno(2.9)$$
which follow by K-conjugation from $L_h$ on $h \subset g$.

\noindent d) Affine-Sugawara nests [14-16].
Repeated K-conjugation on the subgroup
sequence $g \supset h_1 \supset \ldots \supset h_n$ gives the affine-Sugawara
 nests,
$$ \eqalignno{
L_{g/h_1 / \ldots /h_n} & = L_g - L_{h_1/ \ldots /h_n} = L_g + \sum_{j=1}^n
(-)^j
L_{h_j}  &(2.10a) \cr
 c_{g/h_1/ \ldots / h_n} & =c_g - c_{h_1/\ldots /h_n} =c_g + \sum_{j=1}^n (-)^j
c_{h_j} \;\;\;\;.  &(2.10b) \cr} $$
 The nest stress tensors may be rearranged as sums of mutually-commuting
Virasoro constructions for $g/h$ and $h$,
$$ \eqalignno{
 T_{g/h_1/ \ldots  /h_{2n+1} } & = T_{g/h_1} + \sum_{i=1}^{n}
T_{h_{2i}/h_{2i+1}}  &(2.11a) \cr
 T_{g/h_1/ \ldots  /h_{2n} } & = T_{g/h_1} + \sum_{i=1}^{n-1}
T_{h_{2i}/h_{2i+1}} + T_{h_{2n}} &(2.11b) \cr} $$
so the conformal field theories of the affine-Sugawara nests are expected to
be tensor-product field  theories (see Section 5).

\section{Irrational Conformal Field Theory}
Here is an overview of the solution space of the master equation, called
affine-Virasoro space. For further details see the review in Ref.\ci{rd}

\noindent a) Counting [15,18]. The master equation is a set of
$\mbox{dim}\,g (\mbox{dim}\, g + 1)/2$ coupled quadratic equations on the same
number of unknowns $L^{ab}$. This allows us to estimate the number of
inequivalent solutions on each manifold. As an example, there are
approximately $\frac{1}{4}$ billion conformal field theories on each level
of affine $SU(3)$, and exponentially larger numbers on larger manifolds.

\noindent b) Exact solutions
[15,18-26,16].
Large numbers of new solutions have been
found in closed form. On positive integer levels of affine compact $g$, most of
these
solutions are unitary with irrational central charge. As an example, the value
at level 5 of affine $SU(3)$ \ci{hl}
$$ c \left( (SU(3)_5)_{D(1)}^{\#} \right) = 2 \left( 1 - {1 \over \sqrt{61}}
\right) \simeq 1.7439 \eqno(3.1) $$
is the lowest unitary irrational central charge yet observed. See Ref.\ci{lie}
for the most recent list of exact solutions.

\noindent c) Systematics \ci{rd}. Generically, affine-Virasoro space is
organized into level-families of conformal field theories, which are
essentially
analytic functions of the level. On positive integer level of affine compact
$g$, it is clear from the form of the master equation that the generic
level-family has generically irrational central charge.

Moreover, rational central charge is rare in the space of unitary conformal
field theories. Indeed, the rational conformal field theories are rare in the
space of Lie $h$-invariant conformal field theories \ci{lie}, which are
themselves
quite rare. Many candidates for new rational conformal field theories
[23,27,28], beyond the coset constructions, have also been found.

\noindent d) Classification\footnote{In the course of this work, a new
and apparently fundamental connection between Lie groups and graphs was seen.
The interested reader should consult Ref.\ci{lie} and especially Ref.\ci{ggt},
which axiomatizes these observations.}. Study of the space by
high-level\footnote{
The leading behavior $L^{ab} = (P^{ab} /2k) + \cO (k^{-2})$,
$\tilde{L}^{ab} = (\tilde{P}^{ab} /2k) + \cO (k^{-2})$ is believed \ci{hl} to
include all unitary conformal field theories on affine compact $g$, where
$P$ and $\tilde{P}$ are projectors which sum to the inverse Killing metric.
In the graph theories, the projectors are the adjacency matrices of the
graphs.}
expansion \ci{hl} shows a partial classification by graph theory and
generalized graph theories  [18,22-26].
In the classification, each graph is a
level-family, whose conformal field theories carry the symmetry of the graph.
At the present time, seven graph-theory units \ci{rd} of generically unitary
and
irrational conformal field theories have been studied, and it seems likely
that many more can be found \ci{ggt}.

Large as they are, the graph theories cover only very small regions of
affine-Virasoro space. Enough has been learned however to see that all known
exact solutions are special cases of relatively high symmetry, whereas the
generic conformal field theory is completely asymmetric \ci{gt}.

Before going on to the Ward identities, I should mention several other lines
of development.

\noindent 1. Non-chiral CFT's. Adding right-mover copies $\bar{T}$ and
$\tilde{\bar{T}}$ of the K-conjugate stress tensors, we may take the usual
Hamiltonian $H= L^{(0)} + \bar{L}^{(0)}$ for the $L$ theory. Because
the generic CFT has no residual affine symmetry, the physical Hilbert space
of the generic theory $L$ is characterized by
$$ \tilde{L}^{(m>0)} |L \mbox{-physical} \rangle =
\tilde{\bar{L}}^{(m>0)} |L \mbox{-physical} \rangle =0 \;\;\;\;.
\eqno(3.2) $$

\noindent 2. World sheet action \ci{hy2}. Correspondingly, the world-sheet
action  of the generic theory
$L$ is a spin-two gauge theory, in which the WZW theory is gauged by the
K-conjugate theory $\tL$. An open direction here is the relation with
$\sigma$-models and the corresponding space-time geometry of
irrational conformal field theory. In this connection, see also
Ref.\ci{hy1}.

\noindent 3. Superconformal master equations \ci{sme}. The N=1 system has been
studied in some detail [23-26],
and a simplification \ci{jos} of the N=2
system has recently been noted: Eq.(D.5) of Ref.\ci{sme} is redundant, so that
eq.(D.6) is the complete N=2 system. It is an important open problem to find
unitary  N=2 solutions with irrational central charge.

Another open question at N=2 is the relation of the master equation to the
bosonic constructions of Kazama and Suzuki [27,28].

\noindent 4. Exact C-functions [32,30,17]. Exact C-functions \ci{zam}
and associated C-theorems are known for the N=0 and N=1 master equations, but
not yet for N=2.

\section{The Ward Identities of Irrational CFT}
It is clear that the Virasoro master equation is the first step in the study
of irrational conformal field theory (ICFT), which contains rational conformal
field theory (RCFT) as a small subspace of relatively high symmetry,
$$ \mbox{ICFT  } \supset \mbox{ RCFT} \;\;\;\;. \eqno(4.1) $$
On the other hand, the correlators of irrational conformal field theory have,
until recently, remained elusive. The reason is that most of the computational
methods of conformal field theory have been based on null states of extended
Virasoro algebras [5,34-36], whereas the generic conformal field theory,
being totally asymmetric, is not expected to possess such simple algebras.

Recently, Obers and I have found a set of equations, the affine-Virasoro
Ward identities \ci{co}, for the correlators of irrational conformal field
theory.

The first step is to consider KZ-type null states [7,37],
which live in the modules
of the universal covering algebra of the affine algebra, and which are more
general than extended Virasoro null states. As an example, we have
$$ 0 = L^{(-1)} | R_g \rangle - 2 L^{ab} J_a^{(-1)} | R_g \rangle \T_b
\eqno(4.2) $$
which holds for all affine-Virasoro constructions $L^{ab}$. Here
$|R_g \rangle^{\a}, \; \a =1,\ldots ,\mbox{dim}\,\T $ is the affine primary
state with matrix representation $\T$, and the original KZ null state is
recovered from (4.2) by taking $L^{ab}=L_g^{ab} $ and $L^{(-1)}=  L_g^{(-1)}$.
We also choose a so-called $L$-basis of representation $\T$, in which the
conformal weight matrix is diagonal \ci{co}
$$  L^{ab} {(\T_a \T_b)_{\a}}^{\b} = \D_{\a}(\T) \, \d_{\a}^{\b}
\;\;\;\;,\;\;\;\; \a,\b =1 , \ldots ,\mbox{dim}\,\T \;\;\;\; . \eqno(4.3a) $$
$$ L^{m \geq 0} | R_g \rangle =\d_{m,0} \, \D_{\a}(\T)
| R_g \rangle \;\;\;\;,\;\;\;\;
 \tilde{L}^{m \geq 0} | R_g \rangle =\d_{m,0} \,
\tilde{\D}_{\a}(\T)  | R_g \rangle \eqno(4.3b) $$
$$  \D_{g}(\T) =  \D_{\a}(\T) + \tilde{\D}_{\a}(\T) \;\;\;\;. \eqno(4.3c) $$
In such a basis, the states $|R_g \rangle$ are called the broken affine primary
states, because the conformal weights $\D_g(\T)$ of $|R_g \rangle$ under
the affine-Sugawara
construction are broken to the conformal weights $\D_{\a} (\T)$ of the
$L$ theory. These states are also examples of Virasoro biprimary states, which
are simultaneously Virasoro primary under both of the commuting K-conjugate
stress tensors.

To use these null states in correlators, we need something like the
Virasoro primary fields of the $L$ theory. Because we have two commuting
stress tensors, the natural objects are the Virasoro biprimary fields [38,37]
$$ R^{\a}(\bz,z) =\mbox{e}^{(\bz-z) \tL^{(-1)} } R_g^{\a} (z) \,
\mbox{e}^{(z-\bz) \tL^{(-1)}} = \mbox{e}^{(z-\bz) L^{(-1)} } R_g^{\a} (\bz) \,
\mbox{e}^{(\bz-z) L^{(-1)}} \eqno (4.4a) $$
$$ R(z,z) = R_g(z) \;\;\;\;,\;\;\;\;
 R (0,0)  |0 \rangle= | R_g \rangle  \eqno(4.4b) $$
which are  simultaneously Virasoro primary under $T$ and $\tT$.
In (4.4), $R_g^{\a}$ are the broken affine primary fields, and the independent
variable $\bz$ is not necessarily the complex conjugate of $z$. The averages
of these bilocal fields
$$ A^{\a}(\bz,z) =  \langle R^{\a_1}(\T^1,\bz_1,z_1) \ldots
R^{\a_n}(\T^n,\bz_n,z_n) \rangle
\eqno(4.5) $$
are called biconformal correlators.

Inserting the null state (4.2) into the biconformal correlator, the Virasoro
term gives derivatives with respect to the $z$'s, as usual, while the current
term can be evaluated, in terms of the representation matrices, on the
affine-Sugawara line $\bz =z$. More generally, we obtain the
affine-Virasoro Ward identities \ci{co}
$$   \bpa_{j_1} \ldots \bpa_{j_q} \pa_{i_1} \ldots \pa_{i_p} A^{\a}(\bz,z)
 \ve_{\bz=z} = A^{\b}_g (z) { W_{j_1 \ldots j_q, i_1 \ldots i_p }
(z)_{\b}}^{\a} \eqno(4.6) $$
where $W_{ j_1 \ldots j_q ,i_1 \ldots i_p} $ are the affine-Virasoro
connections and $A(z,z) = A_g(z)$ is the affine-Sugawara correlator (which
satisfies the KZ equation on $g$). The first-order connections are
$$  W_{0,i} = 2 L^{ab} \sum_{j\neq i}^n
{\T_a^i \T_b^j \over z_{ij} }  \;\;\;\;,\;\;\;\;\;
W_{i,0} = 2 \tL^{ab} \sum_{j\neq i}^n  { \T_a^i \T_b^j \over z_{ij} }
 \eqno(4.7) $$
which follow from the null state (4.2) and its K-conjugate copy with
$L \ra \tL$. The sum of these two connections is the KZ connection
$W_i^g$, with $L=L_g$.

All the connections may be computed by standard dispersive techniques from the
formula [37,39]
$$  A_g^{\b} (z) {W_{j_1 \ldots j_q, i_1 \ldots i_p } (z)_{\b}}^{\a} =
\;\; \;\;\;\;\;\;\;\;\;\;\; \;\;\;\;\;\;\;\;\;\; $$
$$ \left[ \prod_{r=1}^{q} \tL^{a_r b_r} \oint_{z_{j_r}}
{\mbox{d}\o_r \over 2\p i} \oint_{\o_r}  {\mbox{d}\et_r \over 2\p i} \;
{1 \over \et_r-\o_r} \right] \!
\left[ \prod_{s=1}^{p} L^{c_s d_s} \oint_{z_{i_s}} \!
{\mbox{d}\o_{q+s} \over 2\p i} \oint_{\o_{q+s}} \!
{\mbox{d}\et_{q+s} \over 2\p i} \; {1 \over \et_{q+s}-\o_{q+s}} \right] $$
$$ \times  \langle
J_{a_1}(\et_1)  J_{b_1}(\o_1)  \ldots J_{a_q}(\et_q)  J_{b_q}(\o_q)
J_{c_1}(\et_{q+1})  J_{d_1}(\o_{q+1})  \ldots  $$
$$ \;\;\;\; \;\;\;\;\;\; \; J_{c_p}(\et_{q+p})  J_{d_p}(\o_{q+p})
R_g^{\a_1} (\T^1,z_1) \ldots  R_g^{\a_n} (\T^n,z_n) \rangle
 \eqno(4.8) $$
since the required averages are in the affine-Sugawara theory on $g$. The
results (4.6) and (4.8) prove the existence of the biconformal correlators
(at least
as an expansion about the affine-Sugawara line $\bz =z$), but computation of
all the connections appears to be a formidable task. So far, we have explicitly
evaluated only the first and second-order connections \ci{co} for all theories
and all the connections for the coset constructions [37,39] and
affine-Sugawara nests \ci{wi}.

Many general properties of the connections also follow from (4.8), including
the high-level form of all the connections  \ci{wi}
$$ W_{j_1 \ldots j_q,i_1 \ldots i_p} = W_{j_1 \ldots j_q,0}
W_{0,i_1 \ldots i_p} + \cO (k^{-2})  \eqno(4.9a)$$
$$ W_{j_1 \ldots j_q,0} = \left( \prod_{r=1}^{q-1} \pa_{j_r} \right) W_{j_q,0}
+
\cO (k^{-2}) \;\;\;,\;\; q \geq 1 \eqno(4.9b)$$
$$ W_{0,i_1 \ldots i_p} = \left( \prod_{r=1}^{p-1} \pa_{i_r} \right) W_{0,i_p}
+
\cO (k^{-2}) \;\;\;,\;\; p \geq 1 \eqno(4.9c)$$
and the crossing symmetry of the connections \ci{wi}
$$  W_{j_1 \ldots j_q,i_1 \ldots i_p}  \ve_{k \lra l}
=  W_{j_1 \ldots j_q,i_1 \ldots i_p} \eqno(4.10) $$
where $k \lra l$ includes $T$'s, $z$'s and indices. Moreover,
the consistency \hbox{relations}~\ci{co}
$$ (\pa_i + W_i^g) W_{j_1 \ldots j_q,i_1 \ldots i_p} =
W_{j_1 \ldots j_q i ,i_1 \ldots i_p} + W_{j_1 \ldots j_q,i_1 \ldots i_p i }
\eqno(4.11) $$
show that we need only compute the independent set of connections
$W_{0,i_1 \ldots i_p}$.

So far we have described only the biconformal correlators. To obtain the
conformal correlators, we must factorize the biconformal
correlators\footnote{ The notation in (4.12) includes a sum
$ A^{\a} (\bz,z) = \sum_{\n} ( \bA_{\n} (\bz) \, A_{\n} (z))^{\a} $ over the
conformal structures $\bA_{\n},\; A_{\n}$ of the theories and an assignment
of Lie algebra indices. See Refs.[37,39] and Section 7.}
$$ A^{\a} (\bz,z) = (\bA (\bz)\, A(z) )^{\a} \;\;\;\;,\;\;\;\;
 A_g^{\a} (z) = (\bA (z)\, A(z) )^{\a}
 \eqno(4.12) $$
into the proper correlators $\bA$ and $A$ of the $\tL$ and $L$ theories
respectively. Then the factorized Ward identities \ci{co}
$$   (\pa_{j_1} \ldots \pa_{j_q} \bA \, \pa_{i_1} \ldots \pa_{i_p} A)^{\a}
= (\bA \, A)^{\b}  {( W_{j_1 \ldots j_q, i_1 \ldots i_p } )_{\b}}^{\a}
   \eqno(4.13)$$
are an all-order non-linear differential system for the  correlators of the
K-conjugate pair of  conformal field theories.

\section{Coset and Nest Correlators}
To anchor the new Ward identities, we solved the system first for the rational
conformal field theories.

Given the all-order connections for
$$\tL=L_{g/h} \;\;\;\;,\;\;\;\; L=L_h \eqno(5.1)$$
one solves the
Ward identities (4.6) or (4.13) to obtain the factorized biconformal
correlators [37,39]
$$ A^{\a} (\bz,z)  = A_{g/h}^{\b} (\bz) {A_h(z)_{\b}}^{\a}
\eqno(5.2a) $$
$$A_{g/h}^{\a} (\bz) = A_g^{\b} (\bz) {A_h^{-1}(\bz)_{\b}}^{\a}
 \eqno(5.2b) $$
where $A_{g/h}$ are the coset correlators.  The two-index
symbol $A_h$ is the (invertible) evolution operator of $h$, which solves the
KZ equation for $h\subset g$.

Using the $g$ and $h$-invariant tensors of $\T^1 \otimes \cdots \otimes \T^4$,
one may change to a conformal-block basis for the four-point coset correlators.
Then, the coset blocks [40,37,41]
$$ \C (u)_r{}^R = \F_g(u)_r{}^m \F_h^{-1}(u)_m{}^R \eqno(5.3) $$
are obtained from (5.2b), where $\F_g$ and $\F_h$ are the KZ blocks of $g$
and $h$.

Although the affine-Virasoro Ward identities have provided a derivation
of these coset blocks from first principles, their form was discussed in 1987
by Douglas \ci{do} who argued that they define consistent non-chiral conformal
field theories.

In fact, the coset blocks are the ultimately practical answer for coset
correlators. See \ci{co} for an explicit example on
$(SU(n)_{x_1} \times SU(n)_{x_2}) / SU(n)_{x_1+x_2}$. More generally, the
coset blocks show that coset correlators are always sums of products
of generalized hypergeometric functions (i.e., solutions to the KZ equations),
a conclusion which seems difficult to obtain in other approaches.

Turning next to the simplest affine-Sugawara nest
$$ \tL = L_{g/h_1 /h_2} \;\;\;\;,\;\;\;\; L = L_{h_1 /h_2}\eqno(5.4)$$
one solves the Ward identities
to obtain the biconformal nest correlators \ci{wi}
$$ A^{\a}(\bz,z) = A_{g/h_1}^{\b}(\bz) \, A_{h_1/h_2}(z {)_{\b}}^{\g}
\, A_{h_2}(\bz{)_{\g}}^{\a} \eqno(5.5) $$
where $A_{h_1/h_2} = A_{h_1} A_{h_2}^{-1}$  is composed of the evolution
operators of $h_1$ and $h_2$. The $\bz$ dependence of (5.5) suggests
that the conformal nest correlators are tensor-product theories
$$ A_{g/h_1/h_2} (\bz) = A_{g/h_1}(\bz)  \otimes \, A_{h_2}(\bz) \eqno(5.6) $$
and this has been confirmed \ci{wi} by analysis of (5.5) with the $g,\;h_1$ and
$h_2$-invariant tensors of $\T^1 \otimes \cdots \otimes \T^4$.

Similarly, the biconformal correlators of all the higher nests have been
obtained on  $g \subset h_1 \subset \ldots \subset h_n$, and the conformal nest
correlators $A_{g / h_1/ \ldots /h_n}$ are exactly the tensor-product
theories indicated in the nest stress tensors (2.11). Further details may be
found in Ref.\ci{wi}.

\section{Algebraization of the Ward Identities}
We want to solve the Ward identities for all ICFT, but the differential
system (4.13) is intimidating. In fact, there is an equivalent algebraic
formulation of the system [39], which I will discuss only for four-point
correlators.

To begin, we use the $SL(2) \times SL(2)$ covariance of the biconformal
correlators to
translate our machinery above into invariant form [37]. We obtain the invariant
Ward identities
$$ \bpa^q \pa^p Y^{\a} (\bu,u)\ve_{\bu =u}=Y_g^{\b}(u){W_{qp}(u)_{\b}}^{\a}
 \eqno(6.1a) $$
$$ W_{10} = 2 \tL^{ab} \left( {\T_a^1 \T_b^2 \over u} +
{ \T_a^1 \T_b^3 \over u-1}  \right) \;\;\;\; , \;\;\;\;
 W_{01} = 2 L^{ab} \left( {\T_a^1 \T_b^2 \over u} +
{ \T_a^1 \T_b^3 \over u-1}  \right) \eqno(6.1b) $$
and the invariant factorized Ward identities

$$ (\pa^q \bY \, \pa^p Y)^{\a}  = (\bY\, Y)^{\b} {(W_{qp})_{\b}}^{\a}
\eqno(6.2a) $$
$$ Y^{\a} (\bu,u) = (\bY (\bu)\, Y(u) )^{\a}
\;\;\;\;,\;\;\;\; Y_g^{\a} (u) = ( \bY(u) \, Y(u) )^{\a}  \eqno(6.2b) $$
where $\bu$ and $u$ are cross ratios among the $\bz$'s and $z$'s respectively.
Here, $Y(\bu,u)$ and $Y_g(u)$ are the invariant biconformal correlator and
the invariant affine-Sugawara correlator respectively, while $W_{qp}$ are the
invariant connections  of order $q + p$.

{}From the results for the $n$-point  connections, the invariant connections
are
known for all theories thru second order and to all orders for cosets and
nests.
Similarly, we know the invariant consistency relations
$$ (\pa + W^g) W_{qp} = W_{q+1,p} + W_{q,p+1} \eqno(6.3) $$
the crossing relation
$$  W_{q p} (1-u) = (-)^{q+p}  P_{23} W_{q p} (u) P_{23} \eqno(6.4a) $$
$$ P_{23} \T^2 P_{23} = \T^3 \;\;\;\;,\;\;\;\; P_{23}^2 =1 \eqno(6.4b) $$
and, on simple $g$, we know the high-level form of the invariant connections
$$ W_{q p} =W_{q 0} W_{0 p} + \D_{qp} \;\;\;\;,\;\;\;\;
\D_{qp} =  \cO (k^{-2}) \eqno(6.5a) $$
$$ W_{q 0} = \pa^{q-1} W_{1 0} + \cO (k^{-2}) \;\;\;\,,\;\;\;\;
 W_{0 p} = \pa^{p-1} W_{0 1} + \cO (k^{-2}) \;\;\;\;. \eqno(6.5b)$$
The $q,p$ factorized form seen in (6.5a) is true to all orders with $\D =0$
when
$\tL=L_{g/h}$ and $L=L_h$.

Toward solving the Ward identities, we note first that the
partially-factorized form of the biconformal correlators [39]
$$ Y^{\a}(\bu,u) = \sum_{q,p=0}^{\infty} {(\bu -u_0)^q \over q!}\,
[Y_g^{\b}(u_0) W_{qp} (u_0 {)_{\b}}^{\a} ]\,{(u-u_0)^p \over p!}  \eqno(6.6)$$
solves the unfactorized Ward identities (6.1a), where  $u_0$ is a regular
reference point. The consistency relations (6.3) show that the biconformal
correlators are independent of the choice of $u_0$.

We are left with the problem of factorization, but now we need only factorize
the connections $W_{qp}(u_0)$ at the reference point into functions of
$q$ times functions of $p$. The factorized Ward identities (6.2) have been
reduced to an algebraic problem.

\section{Candidate Correlators for Irrational CFT}
In fact, there are many algebraic factorizations of $W_{qp}$, whose
interrelation is not fully understood. I will discuss here only a
particular solution, which has good physical properties so far as it has been
examined \ci{wi}.

The invariant connections at the reference point define an eigenvalue
problem
$$  \eqalignno{ \sum_p W_{qp} (u_0{)_{\a}}^{\b} \bps_{p \b}^{(\n)} (u_0) & =
E_{\n} (u_0) \bps_{ q \a}^{(\n)} (u_0) &(7.1a) \cr
\sum_q \ps_{q (\n)}^{ \b }(u_0)  W_{qp} (u_0 {)_{\b}}^{\a}  & =
E_{\n} (u_0) \ps_{p(\n) }^{ \a} (u_0)  &(7.1b) \cr} $$
where $\n$, called the conformal structure index, labels the eigenvectors.
Then, the spectral resolution
$$ W_{qp} (u_0 {)_{\a}}^{\b} = \sum_{\n =0}^{\infty}  \bps_{q \a}^{(\n)} (u_0)
E_{\n} (u_0) \, \ps_{p (\n)}^{ \b } (u_0) \eqno(7.2) $$
gives the desired algebraic factorization of the connections, and we obtain
the conformal structures \ci{wi}
$$ Y^{\a} (\bu,u) = (\bY (\bu) \, Y(u) )^{\a}
= \sum_{\n} \bY_{\n} (\bu,u_0) \, Y_{\n}^{\a} (u,u_0) \eqno(7.3a) $$
$$ \bY_{\n} (u,u_0) = \sqrt{E_{\n}(u_0)} \, Y_g^{\a} (u_0)  \bps_{\a}^{(\n)}
(u,u_0)    \;\;\;\;,\;\;\;\; \bps_{\a}^{(\n)} (u,u_0) \equiv\sum_{q=0}^{\infty}
 { (u -u_0)^q \over q!} \,\bps_{q \a}^{(\n)} (u_0) \eqno(7.3b) $$
$$  Y_{\n}^{\a} (u, u_0) = \sqrt{E_{\n}(u_0)} \, \ps_{(\n)}^{\a} (u,u_0)
\;\;\;\;,\;\;\;\;  \ps_{(\n)}^{\a} (u,u_0) \equiv \sum_{p=0}^{\infty}
{ (u-u_0)^p \over p!} \,\ps_{p(\n)}^{ \a } (u_0) \eqno(7.3c) $$
of the $\tL$ and $L$ theories respectively. Because the eigenvalue problem is
infinite dimensional, we find a generically infinite number of conformal
structures for each theory - in accord with intuitive notions about ICFT.

This solution verifies the following properties \ci{wi}.

\noindent 1. Cosets and nests. The solution reproduces the correct coset and
nest correlators above. The mechanism is a {\it degeneracy} of the conformal
structures in which each $\bY_{\n}$ is proportional to the same known
correlators

\noindent 2. Braiding. The solution exhibits a braiding for all ICFT, which
follows from the crossing symmetry (6.4) of the connections and the linearity
of
the eigenvalue problem. Since the coset correlators are correctly included in
the solution, this braiding includes and generalizes the braiding of RCFT.

\noindent 3. Good semi-classical behavior. Because of the factorized high-level
form (6.5) of the connections, we find a similar degeneracy among the
high-level conformal structures of all ICFT. In this way, we identify
the high-level correlators of ICFT,
$$   Y_{\tL}^{\a} (u,u_0)  =  Y_g^{\b} (u_0)  (
\one +  2 \tilde{L}_{ab}  \left[
\T_a^1  \T_b^2 \ln \left( {u \over u_0} \right)
+\T_a^1  \T_b^3 \ln \left( {1-u \over 1-u_0} \right) \right]
 )_{\b}{}^{\a} + \cO (k^{-2})     \eqno(7.4) $$
where  $\tL^{ab} = \tilde{P}^{ab}/ 2k + \cO ( k^{-2})$ is any
affine-Virasoro construction on simple $g$. The result (7.4) correctly includes
the coset and nest correlators, exhibits  physical singularities in all
channels
and shows high-level fusion rules proportional to Clebsch-Gordan coefficients.
These correlators should also be analyzed at the level of conformal blocks.

To further analyze the candidate correlators (7.3), it will be necessary to
know more about the connections. Of particular interest is the next order
in $k^{-1}$, where the high-level degeneracy of the irrational theories is
expected to lift.

\section{Conclusions}
The Virasoro master equation describes irrational conformal field theory, which
includes rational conformal field theory as a small subspace. The
affine-Virasoro Ward identities describe the correlators of irrational
conformal
field theory. The Ward identities are solvable beyond the coset constructions,
and a set of candidate correlators have been obtained for all the irrational
conformal field theories of the master equation.

\section*{Acknowledgement}
I am grateful to N. Obers for his generous help in preparing this manuscript.

\end{document}